\documentclass[twocolumn,showpacs,amssymb,aps, prl]{revtex4}

\begin{document}

\title{The Brown-York mass and the Thorne hoop conjecture}

\author{Niall {\'O} Murchadha}\email{niall@ucc.ie}
\affiliation{Center for Astrophysics, Shanghai Normal University,
 Shanghai, China}
  \affiliation{Physics Department, University College Cork, Cork, Ireland}

\author{Roh-Suan Tung} \email{tung@shnu.edu.cn}
  \affiliation{Center for Astrophysics, Shanghai Normal University,
 Shanghai, China}

\author{Naqing Xie}\email{nqxie@fudan.edu.cn}
  \affiliation{School of Mathematical Sciences, Fudan University, Shanghai, China}
\author{Edward Malec}\email{malec@th.if.uj.edu.pl}
 \affiliation{Physics Department, University College Cork, Cork, Ireland}
  \affiliation{Institute of Physics, Jagiellonian University, Krakow, Poland }

\date{\today}

\begin{abstract}
The Thorne hoop conjecture is an attempt to make precise  the notion that gravitational collapse occurs if enough energy is compressed into a small enough volume, with the `size' being defined by the circumference. We can make a precise statement of this form, in spherical symmetry, using the Brown-York mass as our measure of the energy.
 Consider a spherical 2-surface in a spherically symmetric spacetime. If the Brown-York mass, $M_{BY}$, and the circumference, $C$, satisfy $C < 2\pi M_{BY}$, then the system must either have emerged from a white hole or will collapse into a black hole.  We show that no equivalent result can hold true using either the Liu-Yau mass, $M_{LY}$ or the Wang-Yau mass, $M_{WY}$. This forms a major obstacle to any attempt to establish a Thorne-type hoop theorem in the general case based on either the Liu-Yau or the Wang-Yau mass.
\end{abstract}



\maketitle

It is widely believed that if sufficient energy is concentrated into a small enough volume, the system will gravitationally collapse. Because the concept of the local energy density of the gravitational field is ill-defined, finding a precise version of this statement  has proven very difficult.  All the results which deal with the `interior' are expressed in terms of the `matter' density  only\cite{B}. At the same time, we do know that the gravitational field by itself can cause collapse \cite{BOM}. 

 However, there exist objects which are defined on  closed 2-surfaces in a spacetime which try to quantify the total energy (gravitational and matter) inside the surface. They are called `quasi-local masses'.  There are many quasi-local masses. A comprehensive survey is given by L\'aszl\'o Szabados in \cite{Szabados}. 

Kip Thorne discussed this question of gravitational collapse, avoiding the issue of the interior,  by focusing entirely on the properties of a boundary 2-surface. Thorne's hoop conjecture \cite{Thorne} is

\begin{quote} Horizons form when and only when a mass $M$ gets compacted into a region whose circumference in EVERY direction satisfies $C \lesssim 4 \pi M$. \end{quote}

This statement (deliberately) avoids defining what is meant either by `circumference' or by $M$. When we  deal with spherical symmetry,  it is reasonable to take $C = 2\pi R$, where $R$ is the Schwarzschild or areal radius of the surface in question. This still leaves  open the question of the mass, $M$. Some authors use the mass at spacelike infinity, the ADM mass, but this is clearly unsatisfactory, see \cite {Bo}.  We feel that it is much more natural to use one of the quasi-local masses, which are defined directly for the surface in question. 

In this paper we concentrate on  three closely related quasi-local masses, the Brown-York mass \cite{Brown-York}, $M_{BY}$, the Liu-Yau  mass \cite{L-Y}, $M_{LY}$, and the Wang-Yau mass \cite{Wang-Yau}, $M_{WY}$. These all arise naturally from a Hamiltonian analysis of the Einstein equations. We show that, of the three, only the Brown-York mass allows us to convert the hoop conjecture into a theorem (assuming spherical symmetry); nothing can be proven using the other two. Given that neither the Liu-Yau mass nor the Wang-Yau mass can be defined if the 2-surface is trapped, it is highly unlikely that a hoop-type theorem in the general case can be found using either of these masses.  

This is quite surprising because, in many ways, both the Liu-Yau mass and the Wang-Yau mass are much more `geometric' objects than the Brown-York mass. First, both are intrinsic 2-surface quantities while the Brown-York mass depends on the 3-surface in which the 2-surface is embedded. Second, the Liu-Yau mass is the maximum of the Brown-York mass, whenever both are defined \cite{Niall}. Third, both the Liu-Yau mass and the Wang-Yau mass are always positive, while the Brown-York mass is not. 

We will assume that the spacetime is both asymptotically flat and spherically symmetric. Our computations are made easier because, in spherical symmetry, the Liu-Yau mass equals the Wang-Yau mass. We define $M_{LWY} = M_{LY} = M_{WY}$ in this case. One has some knowledge of the spacetime topology. The spacelike slices could be equivalent to $R^3$, i.e.,  have a `center', or be like $S^2  \times  R$, as in the extended Schwarzschild solution.  If we have two asymptotic ends (at least in a spherical spacetime, with well-behaved matter) we must have both future and past horizons. Therefore we mainly focus our attention on spacetimes with only one end.    An extra, simplifying, assumption is that the spacetime is regular to the past, i.e., that we are only dealing with a non-singular system which may collapse to form a black hole, rather than an object which has emerged from a white hole.  This is the context in which Thorne made his conjecture. 

In particular, we prove:

{\bf Theorem:} Given  a spherically symmetric asymptotically flat spacetime with a regular center and no past singularity and given a spherical 2-surface in it, which is embedded  in a spherical 3-slice, and which satisfies $C < 2\pi M_{BY}$, this surface is trapped.  Further, for a given 2-surface, if $C > 2\pi  M_{BY}$ for all embeddings, the surface is not trapped.

{\bf Countertheorem:} No equivalent theorem holds using the Liu-Wang-Yau mass in a spherically symmetric asymptotically flat spacetime with a regular center and no past singularity. No spherical surface exists which satisfies $C < 2\pi M_{LWY}$. Further, if $C > 2\pi M_{LWY}$ then  this surface can be embedded in a static spacetime with positive matter so  no gravitational collapse need occur.

 Let us begin by introducing some general ideas. There exists a pair of outgoing null rays, $\vec l, \vec m$, at every point on a spacelike 2-surface in a spacetime. We can arrange $\vec l \cdot \vec m = 1/4$. There still remains the freedom to rescale them, i.e., $(\vec l, \vec m) \rightarrow (A\vec l, \vec m /A)$, where $A$ is any positive function on the surface. Associated with $\vec l$ and $\vec m$ are the null expansions, $\rho$, and $\mu$, the fractional rate of change of the 2-area when dragged along the two null normals. Because of the rescaling freedom, we cannot uniquely specify the two null expansions, but the product, $\rho\mu$, is fixed.

If the 2-surface lies in a spacelike 3-slice we have two more normals, one, which we call $\vec v$, is the spacelike normal to the 2-surface in the 3-slice, and the other, call it $\vec u$, is the timelike normal to the 3-slice. Associated with each of them is an expansion. One, the expansion along $\vec v$, is called $k$, and is the 2-mean-curvature of the 2-slice as a surface embedded in the 3-slice. For a round sphere of radius $R$ in flat space we have $k = 2/R$. The expansion along $\vec u$ we call $p$, and it is the 2-trace of the 3-extrinsic curvature of the 3-slice embedded in the spacetime. Given $(\vec u, \vec v)$, there is a natural choice of null normals, $\vec l = (\vec v + \vec u)/\sqrt{8}$ and $\vec m = (\vec v - \vec u)/\sqrt{8}$. This gives an immediate relation between the four expansions, $\rho = (k + p)/\sqrt{8}$ and $\mu = (k - p)/\sqrt{8}$. Therefore we have $8\rho\mu = (k^2 - p^2)$.

The Brown-York energy \cite{Brown-York} is defined for a 2-surface embedded in a spacelike 3-slice. It is given by an integral on the 2-surface
\begin{equation}
E_{BY} = {1 \over 8\pi}\oint(k_0 - k) dA, \label{EBY}
\end{equation}
where $k$ is the mean curvature of the physical embedding and $k_0$ is the mean curvature of the isometric embedding of the 2-surface in a flat 3-space. 

If we restrict our attention to the spherically symmetric case, the energy equals the mass because the linear momentum must be zero.  Let us assume that the spherical surface we consider has area $A$. From this we can work out the Schwarzschild (areal) radius via $4\pi R^2 = A$.  When we isometrically embed this surface in flat space we get a round sphere of area $4\pi R^2$  and radius $R$, with mean curvature $k_0 = 2/R$. Therefore we get
\begin{equation}
{1 \over 8\pi}\oint k_0  dA = {1 \over 8 \pi}\cdot {2 \over R} \cdot 4\pi R^2 = R.
\end{equation}
Hence
 \begin{equation}
M_{BY}  = R -  {1 \over 8\pi}\oint k dA.
\end{equation}
 If the surface satisfies $C = 2\pi R < 2\pi M_{BY}$, then obviously $R < M_{BY}$ and this implies that  the mean curvature of the physical surface, $k$, is negative. Assuming spherical symmetry, the timelike expansion, $p$, is constant on the sphere. If $p$ is positive, $\mu = (k - p)/\sqrt{8} <0$ while if $p$ is negative $\rho = (k + p)/\sqrt{8} < 0$, and if $p = 0$, both $\rho <0$ and $\mu < 0$. Therefore one or other of the null expansions must be negative. If $\mu < 0$ we know that there must be a past singularity.  If we exclude this possibility, we must have $\rho < 0$, and $ \mu > 0$. This is the Penrose definition of a trapped surface \cite{P}.
 
 If the spacelike slice is asymptotically flat, both $\rho$ and $\mu$ are positive near infinity. Therefore there must be an outermost surface on which $\rho$ finally goes positive. This is an apparent horizon. 
 
 If the matter satisfies the null energy condition, there will be an event horizon further out again which  defines a black hole in the future (if $\rho < 0$).   We have that the outermost horizon is the outer limit  of a trapped surface, i.e., we assume $\rho = 0$ while $\mu \geq 0$. Let us write the 4-metric as
 \begin{equation}
 ds^2=-\alpha^2dt^2 +adr^2 +R^2d\Omega^2.
 \end{equation}
 The Raychaudhuri  equation \cite{R}, in spherical symmetry, can be written as
 \begin{eqnarray}
 (\partial_t&+&{\alpha \over \sqrt{a}}\partial_r)\rho = 
({\partial_r\alpha \over \sqrt{a}} + {\alpha \over 2}\sqrt{8}\mu)\rho\nonumber \\
&+& \sqrt{8}\pi \alpha (2j_r/\sqrt{a}- \Delta -   T_r^r)- 8\alpha\rho^2
+\alpha trK\rho , 
 \end{eqnarray}
 where $(\partial_t+{\alpha \over \sqrt{a}}\partial_r)$ is the derivative in the $\vec{l}$ direction, and where 
$(2j_r/\sqrt{a}- \Delta -   T_r^r) \leq 0$ when the null energy condition holds. Let us  start on the apparent horizon with $\rho = 0$ and $\mu \ge 0$ and move in the out-future null direction, i.e., along $\vec{l}$. If we meet matter, $\rho$ goes negative.  This means that the apparent horizon moves outside the local null cone and evolves as a spacelike surface in matter. In vacuum we have that $\rho$ remains equal to zero so the apparent horizon is null. When we finally emerge from the matter, the apparent horizon becomes  a null surface where the spherical cross-sections have constant area. This is the event horizon. 
 
 While $C < 2\pi M_{BY}$ is a sufficient condition that the surface in question be trapped, it is not a necessary condition because of the dependence of the Brown-York mass on the slicing. Rather, one can make the following statement: If the 2-surface is trapped, there exists a spherical 3-slice in which it lies such that $C < 2\pi M_{BY}$. 
  
  Given a trapped surface (and no past singularity) we have that $\rho < 0$ and $\mu >0$. We know that $k = \sqrt{2}(\rho + \mu)$, but  {\it a priori} we have no knowledge of the sign of $k$. However, we can use the scaling freedom. We can multiply $\rho$ by some large number and divide $\mu$ by the same large number. This is equivalent to finding a new spherical 3-slice through the same 2-slice. This will make $k$ as negative as we please. This, in turn,  will make the Brown-York mass as positive as we wish. This implies that if for every embedding we have $C > 2\pi M_{BY}$, the surface cannot be trapped.
 
 Let us consider the equivalent  calculation using either the Liu-Wang-Yau  mass. They are defined in terms of a 2-surface in 4-space. There is no mention of a spacelike 3-slice. They agree in spherical symmetry and are  
 \begin{equation}
 M_{LY} = M_{WY} = M_{LWY} = {1 \over 8\pi}\oint(k_0 - \sqrt{8\rho\mu}) dA, \label{LWY}\end{equation}
 where $k_0$ is again defined as the mean curvature of the isometric embedding in flat 3-space. 
 This gives us
 \begin{equation}
 M_{LWY} =  R - {1 \over 8\pi}\oint \sqrt{8\rho\mu} dA. \end{equation}
 However, this is only well defined if $\rho \mu\ge 0$. Therefore we immediately get
 \begin{equation}
 M_{LY} = M_{WY} = M_{LWY}\leq R. \end{equation}
 
 The condition that $\rho\mu \ge 0$ reduces to two situations, one where both are positive, and one where both are negative. We ignore the situation where both are negative because then we have a singularity to the past as well as to the future.
 We can rescale them to make them equal, $\rho' = \mu' = \sqrt{\rho\mu}$. The associated spacelike and timelike normals are given by $\vec {u}' = \sqrt{2}(\vec {l}' - \vec{m}')$ and $\vec{v}' = \sqrt{2}(\vec{l}' + \vec{m}')$. We get $k' = \sqrt{8\rho\mu} > 0$ and $p' = 0$. This, in turn, means that the spacelike slice defined by $\vec{v}'$ is (locally) a moment of time symmetry slice since $K_{ij} = 0$. Further, relative to this slice we get $M_{BY} = M_{LWY}$. We can  show that this system need not gravitationally collapse, because this 2-surface can be embedded into a static slice of a static spacetime.
 
 This 2-surface can be smoothly joined to an exterior moment-of-time-symmetry slice of the Schwarzschild solution, with mass $m_S$, satisfying
 \begin{equation}
 \sqrt{8\rho\mu} = k' = {2 \over R} {dR\over dL} = {2 \over R}\sqrt{ 1 - {2m_S\over R}}
 \end{equation}
 where $L$ is the proper distance in the radial direction in the slice. This gives
 \begin{equation}
 m_S = R\left({1 \over 2} - R^2\rho\mu\right). \end{equation}
 This is positive because of the following:
 
 {\bf Theorem (Malec - \'O Murchadha)} \cite{MOM}: Given any 2-sphere in any spherically symmetric solution to the Einstein constraints which has a regular center and is asymptotically flat, or has two asymptotic ends, and where the matter source density, $\Delta$, and the current density, $j$, satisfy $\Delta \ge |j|$, then $2R^2\rho\mu \le 1$, and we get equality only at the origin and at infinity.

 {\bf Aside:} This theorem leads to a simple proof  that the Liu-Wang-Yau mass is positive in spherical symmetry.  The bound gives that $\sqrt{8\rho\mu} \le 2/R$, while $k_0 = 2/R$. Therefore the integrand in Eq.(\ref{LWY}) is positive. 
 
 In the interior, we can always choose a thin static shell just inside the surface which is stabilised by a transverse pressure. We can write
 \begin{equation}
 M_{BY}  = M_{LWY} =  R\left(1 - \sqrt{1 - {2m_S \over R}}\right).\label{m_S}
  \end{equation}
  Since the proper matter content of the shell equals the Brown-York mass \cite{Niall}, we get that the surface matter density in the shell, $\sigma$, equals
  \begin{equation}
  \sigma = {1 \over 4\pi R}\left(1 - \sqrt{1 - {2m_S \over R}}\right).
  \end{equation}
  As $2m_S/R \rightarrow 1$, as $\rho\mu \rightarrow 0$, the tangential pressure becomes unboundedly large, it approximates $1/4\pi(R - 2m_S)$. In particular, if $2m_S/R > 48/49$, if $R^2\rho\mu < 1/98$, the stress will be so large that the dominant energy condition is violated \cite{A}. However, this is not a good reason for declaring that this static solution is unphysical. This shell solution, being static, will not gravitationally collapse.

  Note that the formula for the circumference has a $2\pi$ in it rather than a $4\pi$ as in the original Thorne expression. This can be traced back to the fact that the Brown-York mass on the horizon where $R = 2m_S$ equals $2m_S$ (see Eq.(\ref{m_S})), and Thorne was considering something of the order of the Schwarzschild mass.
  
  The key point of this letter is that only the Brown-York mass  makes sense for a trapped region as defined by Penrose. Even if we abandon the special assumption of spherical symmetry neither the Liu-Yau mass nor the Wang-Yau mass is well-defined if we have a trapped surface, i.e., $\rho < 0$ and $\mu > 0$. The definition of the Liu-Yau mass, in the general case,  is as in Eq.(\ref{LWY}), and the square root term makes no sense. The definition of the Wang-Yau mass is more complicated, but, in particular, it requires the mean curvature vector
  \begin{equation}
  h^{\nu} = - kv^{\nu} + pu^{\nu} = -\sqrt{2}(\rho + \mu)v^{\nu} + \sqrt{2}(\rho - \mu)u^{\nu}
  \end{equation}
  be spacelike everywhere on the surface. If the surface satisfies $\rho < 0, \mu > 0$ at any point we know that $|\rho - \mu| > |\rho + \mu|$ so that the mean curvature vector is timelike there.
  
  There is a real possibility that a hoop theorem can be proven using  the Brown-York mass. For example, we have the following result:
  
  {\bf Theorem:} Given a trapped 2-surface, and given any definition of `circumference' which only depends on the 2-geometry of the 2-surface, we can always find a 3-slice so that the Brown-York energy relative to that 3-slice satisfies $C < 2\pi E_{BY}$.
  
  {\bf Proof:} Given that $\rho < 0$, by rescaling we can make $k$ as large and negative as we wish. This will make the Brown-York energy as large and positive as we wish (see Eq.~(\ref{EBY})) . Therefore with a fixed $C$, we can always satisfy the inequality $C < 2\pi E_{BY}$. 
  
  We do not believe that there exists a `best' quasi-local mass. Each quasi-local mass may be a mixture of `good' and `bad' characteristics.  The Brown-York mass has major disadvantages. Consider a spherical 2-surface in Minkowski space. The Liu-Wang-Yau mass vanishes for any such surface. However, the Brown-York mass is negative unless the 3-slice in which it is embedded is flat. Nevertheless, being able to identify those configurations that must gravitationally collapse  is a very physically important attribute and shows the value of the Brown-York mass. It cannot be thought of as an object which  has been superseded  by either the Liu-Yau mass or the Wang-Yau mass.
  
  There are two other quasi-local masses  which arise from the Hamiltonian. Jerzy Kijowski introduced them in an important article \cite{Kijowski}. One he calls `field energy'; this reduces, in spherical symmetry, to the Misner-Sharp-Hawking mass. The other he calls `free energy' and this is identical to the Liu-Yau mass when $\rho\mu \geq 0$. However, both can be defined for trapped surfaces. It would be most interesting to investigate these in further detail.

\begin{acknowledgments}
This work was supported by the
National Natural Science Foundation of China (10771140, 10801036), the
Shanghai Education Development Foundation (05SG45) and NCTS (Taiwan).
N\'OM acknowledges the support of grant 07/RFP/PHYF148. We wish to thank an anonymous referee for very enlightening comments.
\end{acknowledgments}

\end{document}